\documentclass[pra,twocolumn,superscriptaddress,floatfix,showpacs,aps,letterpaper,nobalancelastpage]{revtex4-2}

\usepackage{graphicx} 
\usepackage{color} 
\usepackage{transparent}
\usepackage{amsmath}
\usepackage{hyperref} 
\usepackage{amssymb}
\usepackage{amsthm}
\usepackage{natbib}
\usepackage{qcircuit}
\usepackage{tabulary}

\begin{document}

\title{Native two-qubit gates in fixed-coupling, fixed-frequency transmons beyond cross-resonance interaction}

\date{\today}

 \author{Ken Xuan Wei}
 \email{xkwei@ibm.com}
 \author{Isaac Lauer}
 \author{Emily Pritchett}
 \author{William Shanks}
 \author{David C. McKay}
 \author{Ali Javadi-Abhari}
 \email{ali.javadi@ibm.com}
 \affiliation{IBM Quantum, IBM T.J. Watson Research Center, Yorktown Heights, NY 10598, USA}

\begin{abstract}
Fixed-frequency superconducting qubits demonstrate remarkable success as platforms for stable and scalable quantum computing.
Cross-resonance gates have been the workhorse of fixed-coupling, fixed-frequency superconducting processors, leveraging the entanglement generated by driving one qubit resonantly with a neighbor's frequency to achieve high-fidelity, universal CNOTs. Here, we use on-resonant and off-resonant microwave drives to go beyond cross-resonance, realizing natively interesting two-qubit gates that are not equivalent to CNOTs. In particular, we implement and benchmark native iSWAP, SWAP, $\sqrt{\text{iSWAP}}$, and bSWAP gates; in fact any $SU(4)$ unitaries can be achieved using these techniques. Furthermore, we apply these techniques for an efficient construction of the B-gate: a perfect entangler from which any two-qubit gate can be reached in only two applications. We show these native two-qubit gates are better than their counterparts compiled from cross-resonance gates. We elucidate the resonance conditions required to drive each two-qubit gate and provide a novel frame tracking technique to implement them in Qiskit.
\end{abstract}

\maketitle

\section{Introduction}
Fixed-frequency transmons with fixed-couplings have performed quantum simulations with great success using cross-resonance (CR) gates~\cite{Kim2023nat}, whose fidelities have increased steadily over time~\cite{chow2011,sheldon2016,patterson2019,maple2021}. Recently, fixed-coupling transmons have demonstrated native ZZ interactions using off-resonant Stark tones~\cite{mitchell2021,sizzle2022}. It is natural to ask: can more interesting two-qubit interactions be realized natively in fixed-coupling transmons? In this manuscript we show how to generate native iSWAP, SWAP, bSWAP, and non-Clifford gates such as $\sqrt{\text{iSWAP}}$ and B gates using only microwave drives in a fixed-frequency, fixed-coupling transmon processor. To our knowledge, this is the first experimental demonstration of the B gate, introduced in \cite{Zhang2004}, from which any two-qubit unitary can be compiled with only two applications. These native two-qubit gates are faster and higher fidelity than their counterparts compiled using CZ gates based on the CR interaction. They can be driven in any coupled two-level systems, and do not require higher levels such as the MAP gate~\cite{chow:2013b}. Native two-qubit gates beyond CR can reduce circuit depth for quantum chemistry applications~\cite{googlehf} and improve quantum volume (QV)~\cite{Jurcevic2021} in fixed-coupling superconducting processors. \\

The key challenge in realizing these native two-qubit gates is that the interaction takes place in a frame that is not resonant with either qubit, thus the phases of these two-qubit gates change depending on where they are applied in a quantum circuit. This is drastically different from CR interactions, where the CR tone is applied on the control qubit on-resonant with the target qubit, and thus the CR interaction evolves in the same frame as the target qubit. Analogously, frame-tracking was required in \cite{wei2023} to discover errors in frames that are different than the microwave drive. Here we develop and implement this functionality in Qiskit\cite{Qiskit} by introducing a novel frame tracking technique to correctly compile the time-dependent phases associated with each two-qubit gate in the quantum circuit. We benchmark these native two-qubit gates using interleaved randomized benchmarking (IRB)~\cite{magesan:2011,magesan:2012}. Our manuscript is organized as follows. In \S~\ref{sect:frames} we give an overview of the frame tracking required for implementing phase sensitive two-qubit gates. In \S~\ref{sect:flicforq} we implement and benchmark a high fidelity iSWAP gate by driving one qubit resonantly; combining this with CR allows us to implement a high fidelity direct SWAP gate. In \S~\ref{sect:stark} we show how native iSWAP and bSWAP gates can be implement by driving one qubit off-resonantly. In \S~\ref{sect:nonclif} we implement and benchmark native $\sqrt{\text{iSWAP}}$ and B gates. This work, along with CR and native ZZ, completes all possible native two-qubit interactions for fixed frequency, fixed coupling transmons.
\begin{figure*}[thb!]
\includegraphics[width=1\textwidth]{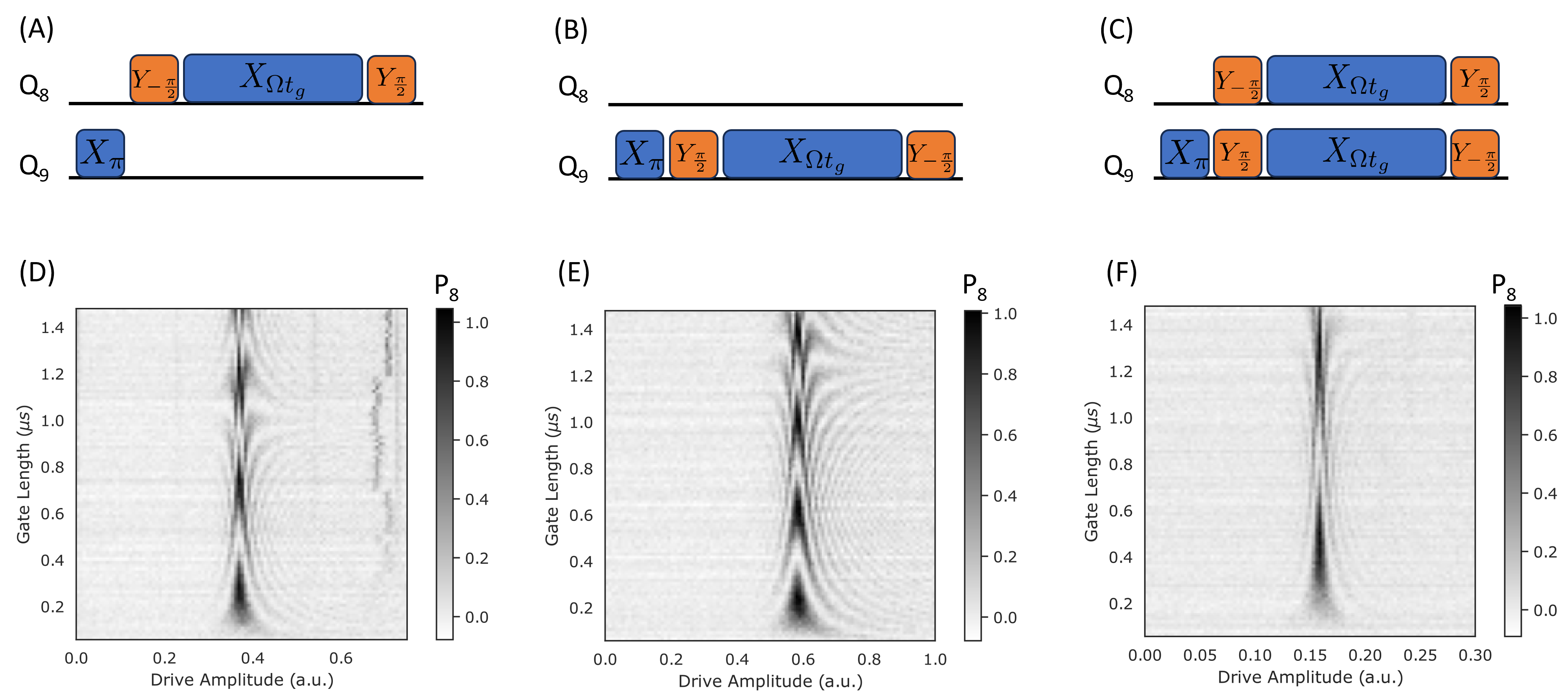}
\caption{Pulse sequences for realizing iSWAP interaction using on-resonant drives. (A) lower frequency qubit $Q_8$ is driven, (B) higher frequency qubit $Q_9$ is driven, and (C) both qubits are driven (FLICFORQ). Coherent oscillations are observed when $\Omega=\Delta$. The excitation probability of $Q_8$, denoted as $P_8$, for the three pulse sequences are shown in (D), (E), and (F). Notice the oscillation rate for one qubit driving is nearly twice the rate as FLICFORQ. For two qubit driving the Rabi amplitudes need not be the same. In (D) we observe TLS interactions characterized by peaks randomly moving in time. Using on-resonant drives (spin-locking) to study TLS dynamics have been investigated in~\cite{leonid2023prxq,abdu2020}.}
\label{flicforq}
\end{figure*}

\section{Frame Tracking
\label{sect:frames}}
In fixed-frequency, fixed-coupling superconducting processors, single qubit and CNOT gates using the CR interaction are implemented in the rotating frame on-resonant with qubit's frequency. Consider two coupled transmons where one qubit is driven either on-resonantly or off-resonantly. The resulting two-qubit unitary $U_g$ is generated by a time-independent Hamiltonian in a frame that is rotating at a different frequency from the driven qubit's resonant frequency by an offset $\Delta$. The unitary implemented in the qubit's frame is given by the following
\begin{align}
\Qcircuit @C=0.5em @R=0.7em{
&\lstick{Q_0}  & \gate{Z(-\Delta t_\text{start})} &  \multigate{1}{U_g} &    \gate{Z(\Delta t_\text{end})}  & \qw  \\
&\lstick{Q_1}  & \qw & \ghost{U_g}   &\qw  &\qw  
}
\label{eq:frame0}
\end{align}
where $Q_0$ and $Q_1$ are the driven and idle qubit respectively, $Z(\theta)$ represents rotation around z by angle $\theta$, and $t_\text{start}$ and $t_\text{end}$ are the start and end times of $U_g$. Here we see that $U_g$ is sandwiched between two time-dependent z-rotations, thus depending on where $U_g$ appears in a quantum circuit, different z-rotations are needed to correct for $Z(-\Delta t_\text{start})$ and $Z(\Delta t_\text{end})$. We refer to adding time-dependent z-rotations to $U_g$ as frame tracking. The situation is simplified for the CR gate, where $Q_0$ is driven at the resonant frequency of $Q_1$. The effective interactions are $ZX$ and $ZI$, both commuting with $Z(\theta)$ on the control qubit. We can move $Z(-\Delta t_\text{start})$ across $U_g$ as shown below
\begin{align*}
\Qcircuit @C=0.5em @R=0.7em{
& \gate{Z(-\Delta t_\text{start})} &  \multigate{1}{U_g} &    \gate{Z(\Delta t_\text{end})}  & \qw  & \raisebox{-2.2em}{$\Rightarrow$} & & \multigate{1}{U_g} &    \gate{Z(\Delta t_g)} & \qw \\
& \qw & \ghost{U_g}   &\qw  &\qw  & & & \ghost{U_g}   & \qw & \qw
}
\end{align*}
where $t_g=t_\text{end}-t_\text{start}$ is the gate time of $U_g$. Thus for two-qubit gates using the CR interaction we only need to apply a time-independent z-rotation at the end, and no frame tracking is needed. We remark that for strongly driven CR gates there are non-Markovian gate errors which prevents $Z(\theta)$ from fully commuting with $U_g$, but these errors can be removed with derivative removal via adiabatic gate (DRAG) pulse~\cite{wei2023,moein2022,chow:2010}. \\

For iSWAP gates, the phases in front of the gate can be moved to the back according to the following rule
\begin{align}
\Qcircuit @C=0.5em @R=0.7em{
& \gate{Z(\phi_1)} &  \multigate{1}{\text{iSWAP}}  & \qw  & \raisebox{-2.2em}{$\Rightarrow$} & & \multigate{1}{\text{iSWAP}} &    \gate{Z(\phi_2)} & \qw\\
& \gate{Z(\phi_2)} & \ghost{\text{iSWAP}}   &\qw  & & & \ghost{\text{iSWAP}}   & \gate{Z(\phi_1)}& \qw
}
\label{eq:iswapfc}
\end{align}
This makes iSWAP compatible with virtual Z rotations~\cite{mckay2017}. The frame tracking for the iSWAP gate will be discussed in detail in \S~\ref{sect:flicforq}. For non-Clifford gates such as $\sqrt{\text{iSWAP}}$, it is no longer possible to move z-rotations across the gate. The frame tracking can be done using physical z rotations, and any single qubit unitary can be implemented as three single qubit gates~\cite{chen2023}. We note that frame tracking has been implemented in tunable architectures such as two fixed-frequency qubits coupled via a tunable qubit~\cite{mckay:2016,tholen2023,ganzhorn2020}, as well as a fixed-frequency qubit coupled to a tunable qubit~\cite{Abrams2020}. \\

\section{On-Resonant direct iSWAP gate
\label{sect:flicforq}}

A direct implementation of iSWAP gate using on-resonant drives has been proposed before in~\cite{rigetti2005}; to the best of our knowledge, these gates have not been experimentally implemented and benchmarked. The mechanism, called Fixed LIn-ear Couplings between Fixed Off-Resonant Qubits (FLICFORQ), is based on driving two coupled-qubits resonantly satisfying the resonance condition $\Omega_{R1}+\Omega_{R2}=|\Delta_{12}|$, where $\Omega_{R1}$ and $\Omega_{R2}$ are the Rabi amplitudes of the resonant drives and $\Delta_{12}$ is the frequency difference between the qubits. We implement a similar approach as FLICFORQ, but we drive only on one qubit resonantly. This has two advantages: first we observe that the iSWAP rate for one qubit driving is nearly twice the rate for dual qubit driving (FLICFORQ). In addition, it frees up the other qubit, which we can drive separately to achieve a native SWAP as well. The pulse sequences and iSWAP dynamics are presented in Fig.~\ref{flicforq}; the oscillations at the resonance points and the rates are provided in Appendix~\ref{sect:osci}.\\

\begin{figure}[t!]
\includegraphics[width=0.45\textwidth]{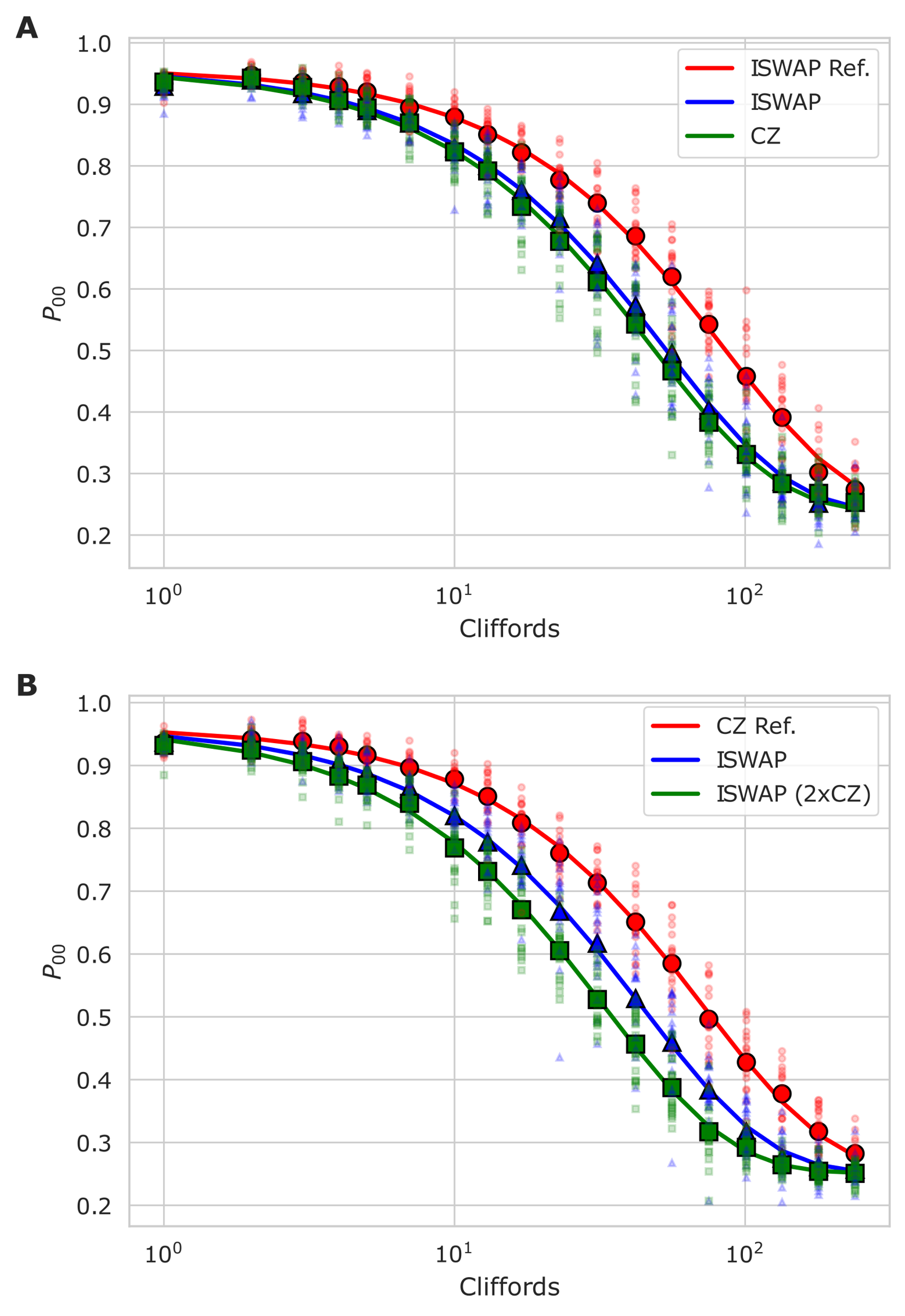}
\caption{Interleaved RB data. {\textbf{A}} uses native iSWAP in the reference RB, and interleaves native iSWAP and CZ based on CR. {\textbf{B}} uses CZ gate in the reference RB, and interleaves native iSWAP and compiled iSWAP using two CZ. The native iSWAP gate is realized by driving $Q_8$ resonantly, and CZ gate is realized by driving $Q_9$ at the frequency of $Q_8$ (CR). $P_{00}$ represent the probability of returning to $|00\rangle$ state. Large plot markers represent the average over 20 RB realizations (small markers); solid lines are exponential fits to the average. The measured EPG for the native iSWAP gate is $0.00648(21)$.}
\label{iswap}
\end{figure}

The starting point of our analysis is the Hamiltonian of an on-resonant drive along the x-axis, after the rotating wave approximation (RWA) the effective Hamiltonian is given by~\cite{wei2023,magesan:2020}
\begin{table}[b!]
\begin{ruledtabular}
	\begin{tabular}{c c c c}
	Parameters &  $Q_8$  &  $ Q_9$ & $Q_{12}$\\
	\hline \vspace{0.0005\linewidth} \\
	$f_{01} \ ({\rm GHz})$ &  $4.726$  & 4.780 & 4.850 \\
	$\alpha$ ({\rm MHz}) &   $-333$  & -332 & -332 \\
	$T_1$ \ ($\mu$s) &   $277(9)$  & 231(5) & 184(4)\\
	$T_{2\rm echo}$\ ($\mu$s) &   $53(4)$  & 219(14) & 334(26)\\
	1Q RB &   $2.24(53)$e$-4$ & $1.50(36)$e$-4$ & $1.69(40)$e$-4$   \\
	\end{tabular}
\end{ruledtabular}
\caption{Summary of qubit parameters used in this work. The 1Q RB numbers are error per Cliffords (EPC) measured from the gate set consists of $X_{\pm \pi/2}$, $Y_{\pm \pi/2}$, $Z_{\pm \pi/2}$, and $Z_{0,\pi}$; all $Z$ gates are virtual frame changes~\cite{mckay2017}. On the quantum processor, $Q_{8,9}$ are connected and $Q_{8,12}$ are connected.}
\label{qubitparams}
\end{table}
\begin{align*}
    H_{D}=\frac{\Omega}{2} (XI + \mu ZX) -\frac{\Delta}{2} IZ
\end{align*}
where $\Omega$ is the Rabi amplitude of the drive, $\Delta$ is the frequency difference between the driven and idle qubit, and $\mu=J/\Delta$ with $J\approx 2$ MHz being the coupling rate between the two qubits. $\Omega$ is always taken as positive, and for simplicity we have ignored small cross-talk terms $IX$ and $ZZ$. Typically for single qubit gates we operate in the regime $\Omega \ll |\Delta|$, where spectator error arising from the CR ($ZX$) term is small and can be neglected. As we increase the driving amplitude, significant spectator error can be observed~\cite{wei2023}. For $\Omega=|\Delta|$, the spectator error becomes the dominant interaction and it can be used to drive an iSWAP gate. To construct the iSWAP gate, we sandwich $H_{D}$ between single qubit rotations $Y_{\pm\pi/2}$ on the driven qubit and calibrate $\Omega$ and gate time $t_g$. The effective Hamiltonian becomes
\begin{align*}
    Y_{\pm \pi/2}H_{D}Y_{\mp \pi/2}=\mp \frac{\Omega}{2} (ZI - \mu XX) -\frac{\Delta}{2} IZ \\
    =\mp \frac{\Omega}{2}ZI - \frac{\Delta}{2}IZ \pm \frac{\mu\Omega}{4}(XX+YY)\pm \frac{\mu\Omega}{4}(XX-YY)
\end{align*}
Depending on the sign of $\Delta$, one can choose the appropriate single qubit rotations $Y_{\pm\pi/2}$ to drive either iSWAP ($XX+YY$) or bSWAP ($XX-YY$). In this section we focus on iSWAP. In section~\ref{sect:stark} we show that iSWAP and bSWAP can also be realized from off-resonant drives. The final unitary needs frame tracking to become iSWAP. Using Eq.~(\ref{eq:frame0}) and Eq.~(\ref{eq:iswapfc}), we can parameterize the unitary as the following,
\begin{align}
&\Qcircuit @C=0.5em @R=0.7em{
 &  \multigate{1}{\text{iSWAP}}  & \qw  & \raisebox{-2.2em}{$=$} & & \multigate{1}{U_g} &    \gate{Z(\Delta t_\text{end} + \phi_1)}& \qw \\
 & \ghost{\text{iSWAP}}   &\qw  & & & \ghost{U_g}   & \gate{Z(-\Delta t_\text{end} + \phi_2)}& \qw
} \\
&\Qcircuit @C=0.5em @R=0.7em{
 & \raisebox{-2.2em}{$=$} & & \gate{Y_{\pi/2}} & \gate{X_{\Omega t_g}(\Omega=\Delta)} & \gate{Y_{-\pi/2}} &    \gate{Z(\Delta t_\text{end} + \phi_1)}& \qw \\
 & & & \qw & \qw & \qw   & \gate{Z(-\Delta t_\text{end} + \phi_2)}& \qw
} \nonumber
\label{eq:frameiswap}
\end{align}
where $U_g=Y_{\pi/2}X_{\Omega t_g}Y_{-\pi/2}$ and $X_{\Omega t_g}=e^{-i H_D t_g}$. $\phi_1$ and $\phi_2$ are time-independent phases; their calibration sequences are presented in Appendix~\ref{sect:iswapcal}. In order to use this iSWAP gate in a quantum circuit, we construct a pass manager in Qiskit which does the following: (1) schedule the quantum circuit, locate each iSWAP gate and insert two virtual Z-gates (frame changes) at the end of iSWAP gate according to Eq.~(\ref{eq:frameiswap}); (2) move all the frame changes to the end of the circuit, move frame changes across iSWAP gates according to Eq.~(\ref{eq:iswapfc}). Note that z rotations remain virtual throughout the circuit and contribute no extra time, thus preserving the validity of the original schedule. A similar approach has been demonstrated recently to track iSWAP frames in a Fluxonium processor~\cite{bao2022}. We apply the aforementioned pass manager to interleaved randomized benchmarking (IRB) circuits to benchmark native iSWAP gates constructed from on-resonant drives. The IRB data are shown in Fig.~\ref{iswap}. The native iSWAP gate is realized by resonantly driving the lower frequency qubit $Q_8$. We measure an error per gate (EPG) of $0.00648(21)$ for the native iSWAP gate, an EPG of $0.01224(27)$ for the compiled iSWAP using two CZ, and an EPG of $0.00666(24)$ for CZ. For the IRB experiments, we use 64 ns single qubit gates and 270 ns on-resonant drive so the total iSWAP gate time is 398 ns. The CZ gate is an echoed CR gate with target rotary~\cite{sheldon2016,Jurcevic2021,sundaresan2020prxq} and has a gate time of 512 ns. The pulse sequence for the CZ gate is shown in Appendix~\ref{sect:cz}. The single qubit gates are Gaussian pulses with 4$\sigma$ long. Both the on-resonant and CR drives are flat-topped Gaussian pulses where rise and fall are $2\sigma$ long with $\sigma\approx 14.22$ ns. All qubit parameters for the experiments including coherence times are summarized in Table.~\ref{qubitparams}; device properties of the quantum processor can be found in the supplemental materials in~\cite{Kim2023nat}. We also note a recent work implementing iSWAP using simultaneous CR interactions~\cite{heya2023}.\\

\subsection{Direct SWAP gate}
\begin{figure}[t!]
\includegraphics[width=0.45\textwidth]{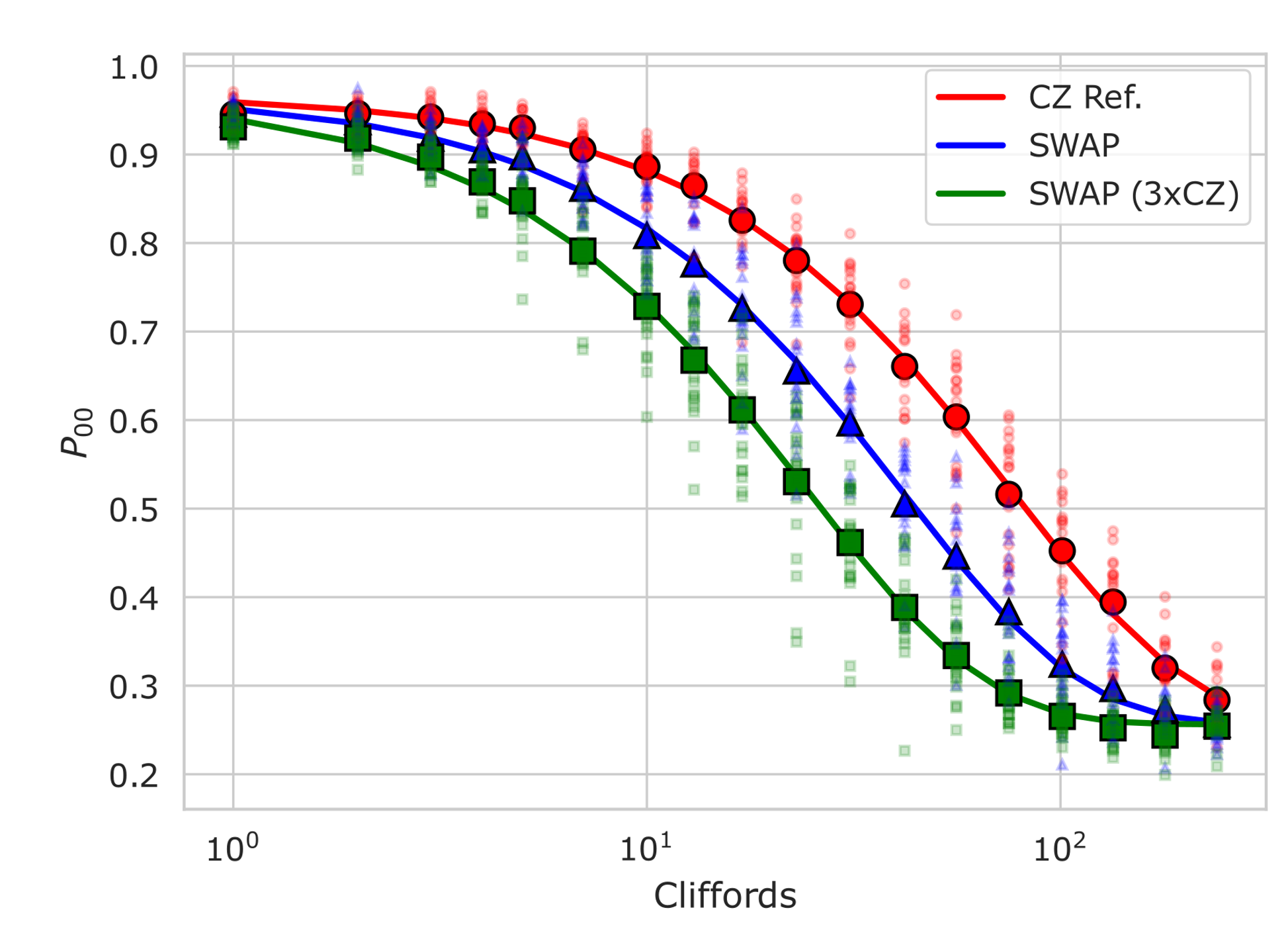}
\caption{Interleaved RB data for native SWAP gate realized by simultaneously driving $Q_8$ on-resonant and $Q_9$ off-resonant. The reference RB uses CZ gate based on CR; the interleaved gates are native SWAP and compiled SWAP using three CZ gates. The measured EPG for the native SWAP gate is $0.00821(23)$.}
\label{swap}
\end{figure}

We can augment the direct iSWAP gate constructed using on-resonant drives into a direct SWAP gate by applying a CR pulse on the idling qubit. In addition to $XX+YY$, SWAP gate needs $ZZ$. The CR pulse on the idling qubit generates $XZ$, which gets transformed into $ZZ$ by the single qubit $Y_{\pm \pi/2}$ rotations. The frame tracking for the SWAP gate is identical to the iSWAP gate, as shown below
\begin{align}
&\Qcircuit @C=0.5em @R=0.7em{
 &  \multigate{1}{\text{SWAP}}  & \qw  & \raisebox{-2.2em}{$=$} & & \multigate{1}{U_g} &    \gate{Z(\Delta t_\text{end} + \phi_1)}& \qw \\
 & \ghost{\text{SWAP}}   &\qw  & & & \ghost{U_g}   & \gate{Z(-\Delta t_\text{end} + \phi_2)}& \qw
} \\
&\Qcircuit @C=0.4em @R=0.7em{
 & \raisebox{-2.2em}{$=$} & & \gate{Y_{\pi/2}} & \gate{X_{\Omega t_g}(\Omega=\Delta+\omega_\text{s})} & \gate{Y_{-\pi/2}} &    \gate{Z(\Delta t_\text{end} + \phi_1)}& \qw \\
 & & & \qw & \gate{\text{Cross Resonance}} & \qw   & \gate{Z(-\Delta t_\text{end} + \phi_2)}& \qw
}\nonumber
\end{align}
The phases $\phi_{1,2}$ are calibrated differently from iSWAP gate, see Appendix~\ref{sect:swapcal} for details. The calibration routine for the direct SWAP gate is the following. First the amplitude of CR drive is calibrated to generate a $XZ$ rotation of $\pi/2$ for the same gate time as the direct iSWAP gate. Then the Rabi amplitude $\Omega$ of the on-resonant is calibrated, notice that the resonance condition is $\Omega=\Delta+\omega_\text{s}$ where $\omega_\text{s}$ is the Stark shift on the idling qubit generated by the CR pulse. Finally $\phi_{1,2}$ are calibrated and the frame tracking is applied in software via the pass manager in Qiskit. We show benchmarking results for the direct SWAP gate in Fig.~\ref{swap}. The direct SWAP gate uses the same set of qubits ($Q_{8,9}$), and has the same length as the iSWAP gate (398 ns). We realize the SWAP gate by driving resonantly on $Q_8$ and simultaneously applying CR on $Q_9$ at $Q_8$'s frequency. The measured EPG are $0.00821(23)$ for native SWAP gate and $0.02039(38)$ for compiled SWAP using three CZ.

\section{Off-resonant (Stark) iSWAP and bSWAP gates
\label{sect:stark}}

\begin{figure}[t!]
\includegraphics[width=0.45\textwidth]{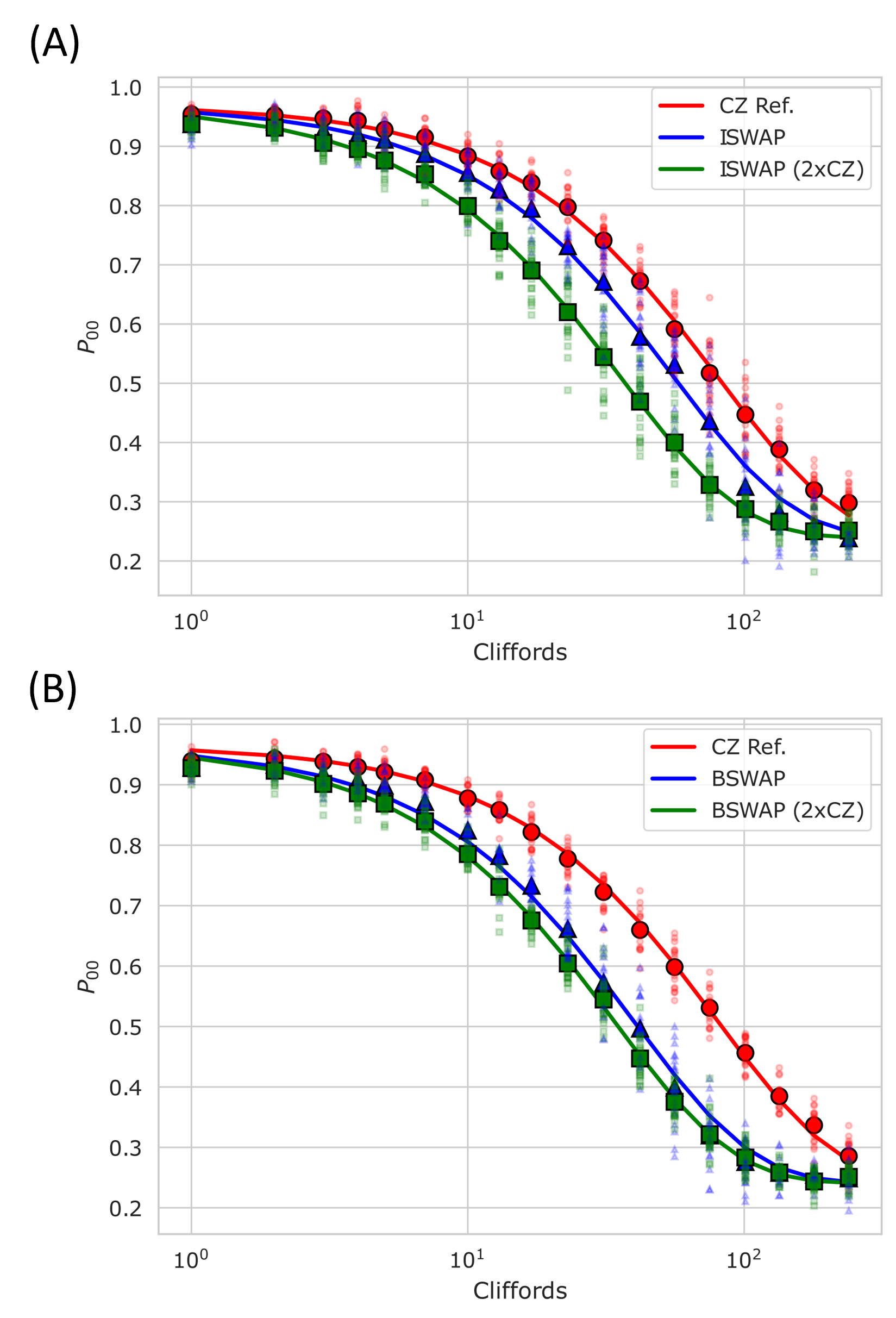}
\caption{Interleaved RB data for (A) Stark iSWAP and (B) bSWAP gates. The IRB experiments used 20 realizations. The measured EPG for the Stark iSWAP gate is $0.00409(27)$ and for the Stark bSWAP gate is $0.00936(39)$.}
\label{stark}
\end{figure}

A recent work shows that iSWAP gate can be realized by driving both qubits off-resonantly~\cite{nguyen2022}. The idea is to drive the lower frequency qubit from below and the higher frequency qubit from above, iSWAP is realized when the sum of the Stark shifts becomes equal to the detuning of the qubits. By adding a third tone to one of the qubit, it is possible to realize programmable Heisenberg interactions~\cite{nguyen2022}. Here we generate a native iSWAP gate using one off-resonant drive with the resonance condition $\omega_\text{s}=-\Delta$. To see how the gate works, we start from following drive Hamiltonian
\begin{align*}
    H_{D}=\frac{\Omega}{2} (XI + \mu ZX) +\frac{\lambda}{2} ZI-\frac{\Delta-\lambda}{2} IZ
\end{align*}
where $\lambda=\omega_1-\omega_D$ is the detuning of the drive and $\Delta$ is the frequency difference between the two qubits. We can rewrite the drive Hamiltonian as 
\begin{align*}
H_{D}=\frac{\Omega'}{2} Z'I-\frac{\Delta-\lambda}{2} IZ + \frac{\mu \Omega}{2} ZX 
\end{align*}
where $\Omega'=\sqrt{\Omega^2+\lambda^2}$ and $Z'=(\Omega X + \lambda Z)/\Omega'$. Since  the axes $Z'$ and $Z$ are not the same, the Stark iSWAP gate will have non-Markovian gate errors which must be corrected by applying DRAG~\cite{wei2023,nguyen2022}. Intuitively, DRAG correction works by rotating $Z'$ into $Z$, its function is similar to the $Y_{\pm \pi/2}$ gates in on-resonant iSWAP. The final Hamiltonian for the Stark iSWAP is given by
\begin{align}
H_{D}=\frac{\text{sign}(\lambda)\Omega'}{2} ZI-\frac{\Delta-\lambda}{2} IZ - \frac{\mu \Omega\lambda}{2\Omega'} XX + \frac{\mu \Omega^2}{2\Omega'} ZX 
    \label{eq:drive}
\end{align}
In order to drive iSWAP, we need the coefficients of $ZI$ to match that of $IZ$. This leads to the resonance condition $\omega_\text{s}=-\Delta$, where $\omega_\text{s}$ is the Stark shift defined as $\omega_\text{s}=\text{sign}(\lambda)\sqrt{\Omega^2+\lambda^2} - \lambda$. The frame tracking shown below for the Stark iSWAP gate is exactly the same as the native iSWAP gate realized with on-resonant drive. The only difference being that the single qubit rotations $Y_{\pm \pi/2}$ are no longer needed. 
\begin{align*}
&\Qcircuit @C=0.5em @R=0.7em{
 &  \multigate{1}{\text{iSWAP}}  & \qw  & \raisebox{-2.2em}{$=$} & & \multigate{1}{U_g} &    \gate{Z(\Delta t_\text{end} + \phi_1)}& \qw \\
 & \ghost{\text{iSWAP}}   &\qw  & & & \ghost{U_g}   & \gate{Z(-\Delta t_\text{end} + \phi_2)}& \qw
} \\
&\Qcircuit @C=0.5em @R=0.7em{
 & \raisebox{-2.2em}{$=$} & & \gate{\text{Off Resonant Drive}}  &    \gate{Z(\Delta t_\text{end} + \phi_1)}& \qw \\
 & & & \qw   & \gate{Z(-\Delta t_\text{end} + \phi_2)}& \qw
}
\end{align*}
We show IRB results in Fig.~\ref{stark}(A). The Stark iSWAP gate is realized by driving $Q_8$ 60 MHz below its resonant frequency. The measured EPG for the Stark iSWAP gate is $0.00409(27)$, and the gate time is 192 ns. We remark that we can also augment the Stark iSWAP gate into a native SWAP gate by driving the idling qubit off-resonantly at the same frequency, and adjust the phase differences between the two drives to get $ZZ$ in addition to $XX+YY$ using the mechanism described in~\cite{mitchell2021,sizzle2022,nguyen2022}. 

Next we show how to realize a native bSWAP gate by driving one qubit off-resonantly. The bSWAP has the interaction $XX-YY$ and is unitarily equivalent to iSWAP up to single-qubit gates~\cite{poletto2012}. From the drive Hamiltonian in Eq.~(\ref{eq:drive}), we can resonantly excite $XX-YY$ provided that the coefficients of $ZI$ is negative of that of $IZ$, this leads to the resonance condition $\omega_1 + \omega_2 + \omega_\text{s}=2\omega_D$. Here $\omega_{1,2,\text{s},D}$ are the qubit one frequency, qubit two frequency, Stark shift, and drive frequency respectively. Phases can be moved across bSWAP in the following manner
\begin{align*}
\Qcircuit @C=0.5em @R=0.7em{
& \gate{Z(\phi_1)} &  \multigate{1}{\text{bSWAP}}  & \qw  & \raisebox{-2.2em}{$\Rightarrow$} & & \multigate{1}{\text{bSWAP}} &    \gate{Z(-\phi_2)} \\
& \gate{Z(\phi_2)} & \ghost{\text{bSWAP}}   &\qw  & & & \ghost{\text{bSWAP}}   & \gate{Z(-\phi_1)}
}
\end{align*}
Combined with the following frame tracking
\begin{align*}
&\Qcircuit @C=0.5em @R=0.7em{
 &  \multigate{1}{\text{bSWAP}}  & \qw  & \raisebox{-2.2em}{$=$} & & \multigate{1}{U_g} &    \gate{Z(\omega_\text{s} t_\text{end} + \phi_1)}& \qw \\
 & \ghost{\text{bSWAP}}   &\qw  & & & \ghost{U_g}   & \gate{Z(\omega_\text{s} t_\text{end} + \phi_2)}& \qw
} \\
&\Qcircuit @C=0.5em @R=0.7em{
 & \raisebox{-2.2em}{$=$} & & \gate{\text{Off Resonant Drive}}  &    \gate{Z(\omega_\text{s} t_\text{end} + \phi_1)}& \qw \\
 & & & \qw   & \gate{Z(\omega_\text{s} t_\text{end} + \phi_2)}& \qw
}
\end{align*}
we can benchmark bSWAP in Qiskit just the same as iSWAP. However bSWAP frame tracking uses $\omega_\text{s}$ instead of the detuning between qubits $\Delta$. The phases for the bSWAP gate are calibrated according to Appendix~\ref{sect:bswapcal}. We realize native bSWAP gate for two qubits ($Q_{8,12}$) separated by 124 MHz by driving the higher frequency qubit 44 MHz below its resonant frequency. The bSWAP gate has a gate time of 462 ns. In Fig.~\ref{stark}(B) we show IRB result of bSWAP. We measure an EPG of $0.00936(39)$ for native bSWAP and $0.01265(46)$ for compiled bSWAP using two CZ. \\

We expect bSWAP gate fidelity to decrease for qubit pairs with small detunings, since the driving frequency will become too close to the qubit frequency. For large detuning pairs, we expect iSWAP gate fidelity to decrease because of the large Rabi drive and Stark shift required. Thus, having both bSWAP and iSWAP available in a multi-qubit processor can be beneficial as they can complement each other. Finally, we expect the last term in Eq.~(\ref{eq:drive}) to have a small contribution to the gate error when resonance conditions are met.

\section{Non-Clifford gates
\label{sect:nonclif}}

\subsection{$\sqrt{\text{iSWAP}}$ gate}

We present two realizations of $\sqrt{\text{iSWAP}}$ gate using on-resonant and off-resonant drive on one qubit. There are two crucial differences between realizing iSWAP gate and $\sqrt{\text{iSWAP}}$ gate. Since we can no longer move frame changes across the gate like we did for iSWAP, frame tracking must be done before and after the gate. Secondly, we need three instead of two time-independent phases to tune up the $\sqrt{\text{iSWAP}}$ gate. The frame tracking is done according to the following
\begin{align*}
&\Qcircuit @C=0.4em @R=0.7em{
 &  \multigate{1}{\sqrt{\text{iSWAP}}}  & \qw  & \raisebox{-2.2em}{$=$} & & \gate{Z(-\Delta t_\text{end}+\phi_1)}  & \multigate{1}{U_g} &    \gate{Z(\Delta t_\text{end} + \phi_2)}& \qw \\
 & \ghost{\sqrt{\text{iSWAP}}}   &\qw  & & & \qw & \ghost{U_g}   & \gate{Z(\phi_3)}& \qw
}
\end{align*}

\begin{figure}[t!]
\includegraphics[width=0.45\textwidth]{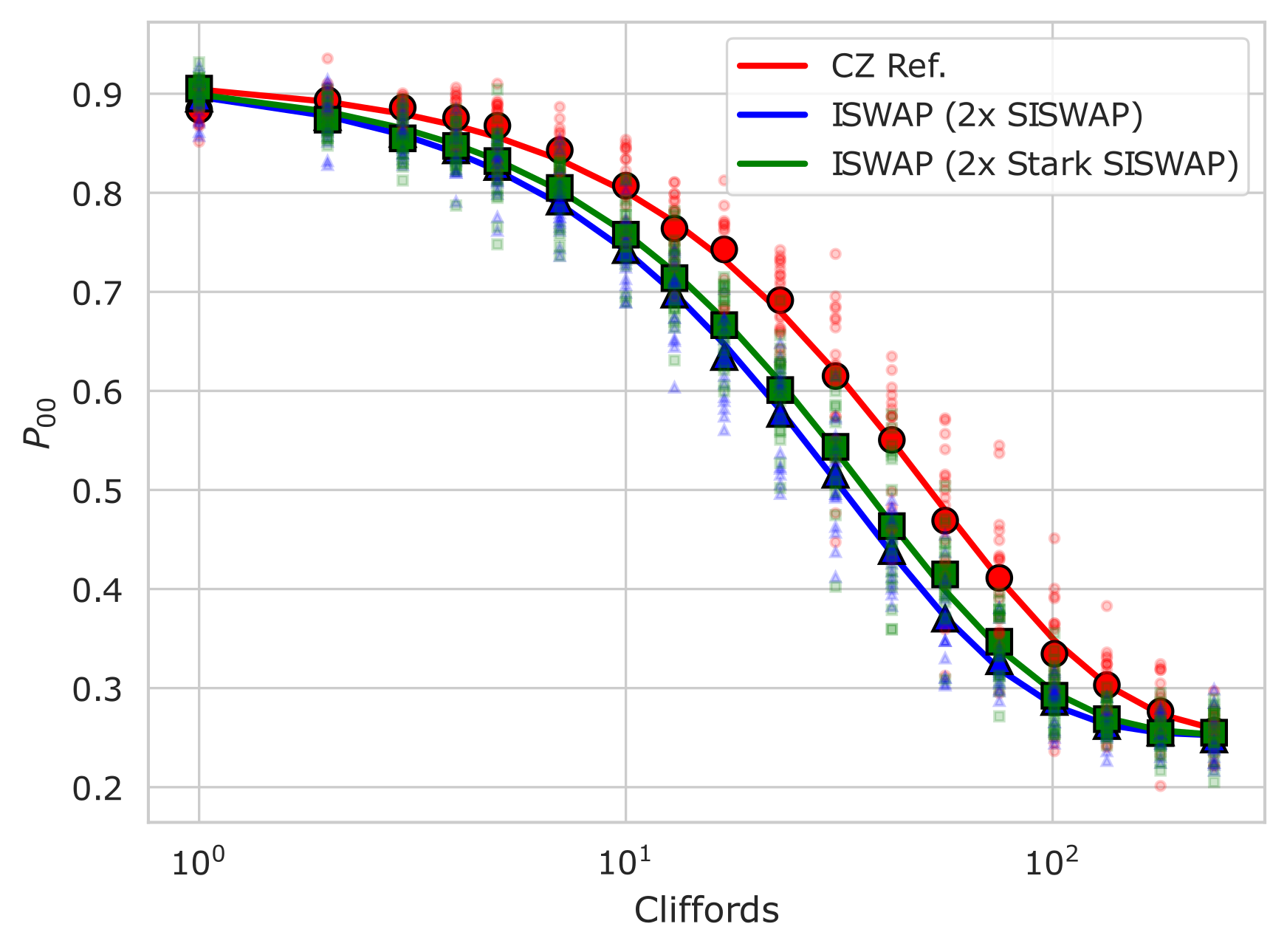}
\caption{Interleaved RB data for $\sqrt{\text{iSWAP}}$ gate. 20 RB realizations are used. The interleaved element is an iSWAP gate composed of two $\sqrt{\text{iSWAP}}$ gates. The EPG are estimated by dividing the EPG of iSWAP by two. The EPG for the on-resonant $\sqrt{\text{iSWAP}}$ gate is $0.00427(16)$, and the EPG for the Stark $\sqrt{\text{iSWAP}}$ gate is $0.00294(15)$.}
\label{sqiswap}
\end{figure}

where $U_g$ is realized by driving the first qubit either on-resonant sandwiched by single qubit gates or off-resonant. We need to calibrate three instead of four time-independent phases because of the property $[XX+YY,ZI+IZ]=0$; we can always move the same phases on both qubits across $\sqrt{\text{iSWAP}}$. The procedure for calibrating $\phi_{1,2,3}$ are outlined in Appendix~\ref{sect:siswapcal}. 

For direct $\sqrt{\text{iSWAP}}$ gate using on-resonant drives, we have modified the pass manager to first combine the two single qubit gates $Y_{\pm\pi/2}$ with other single qubit gates between two $\sqrt{\text{iSWAP}}$ gates, then apply the frame tracking. This allows us to further shorten the gate time. For direct $\sqrt{\text{iSWAP}}$ gate using off-resonant drives, we reduce the length of the Stark iSWAP gate to approximately half. The total gate time for the on-resonant $\sqrt{\text{iSWAP}}$ is 149 ns and for the off-resonant $\sqrt{\text{iSWAP}}$ is 117 ns. We use IRB to benchmark iSWAP made out of two direct $\sqrt{\text{iSWAP}}$ gates, the results are shown Fig.~\ref{sqiswap}. Note that physical z rotations no longer preserve the original schedule, and thus the act of frame tracking itself changes the time-dependent frames in the circuit. Therefore we provision extra general single-qubit gates (U) prior to scheduling. All frames as well as any inherent single-qubit gates of the circuit will be absorbed into this U gate, which makes the schedule deterministic. All U gates are constructed using three pulses~\cite{chen2023}. 

We observe improved EPG for $\sqrt{\text{iSWAP}}$ over iSWAP gates for both on-resonant and off-resonant driving. While $\sqrt{\text{iSWAP}}$ has lower error rates and is more efficient than iSWAP in constructing general SU(4) operations, its inability to work with virtual z-gates may undermine the aforementioned advantages in processors where single qubit gates are less ideal. The technique used to realize Stark $\sqrt{\text{iSWAP}}$ can be applied to realize $\sqrt{\text{bSWAP}}$ gate as well. We note recent works have also realized $\sqrt{\text{iSWAP}}$ in Fluxonium processors~\cite{huang2023, zhang2023tunable}. \\

\subsection{B gate}

\begin{figure}[t!]
\includegraphics[width=0.45\textwidth]{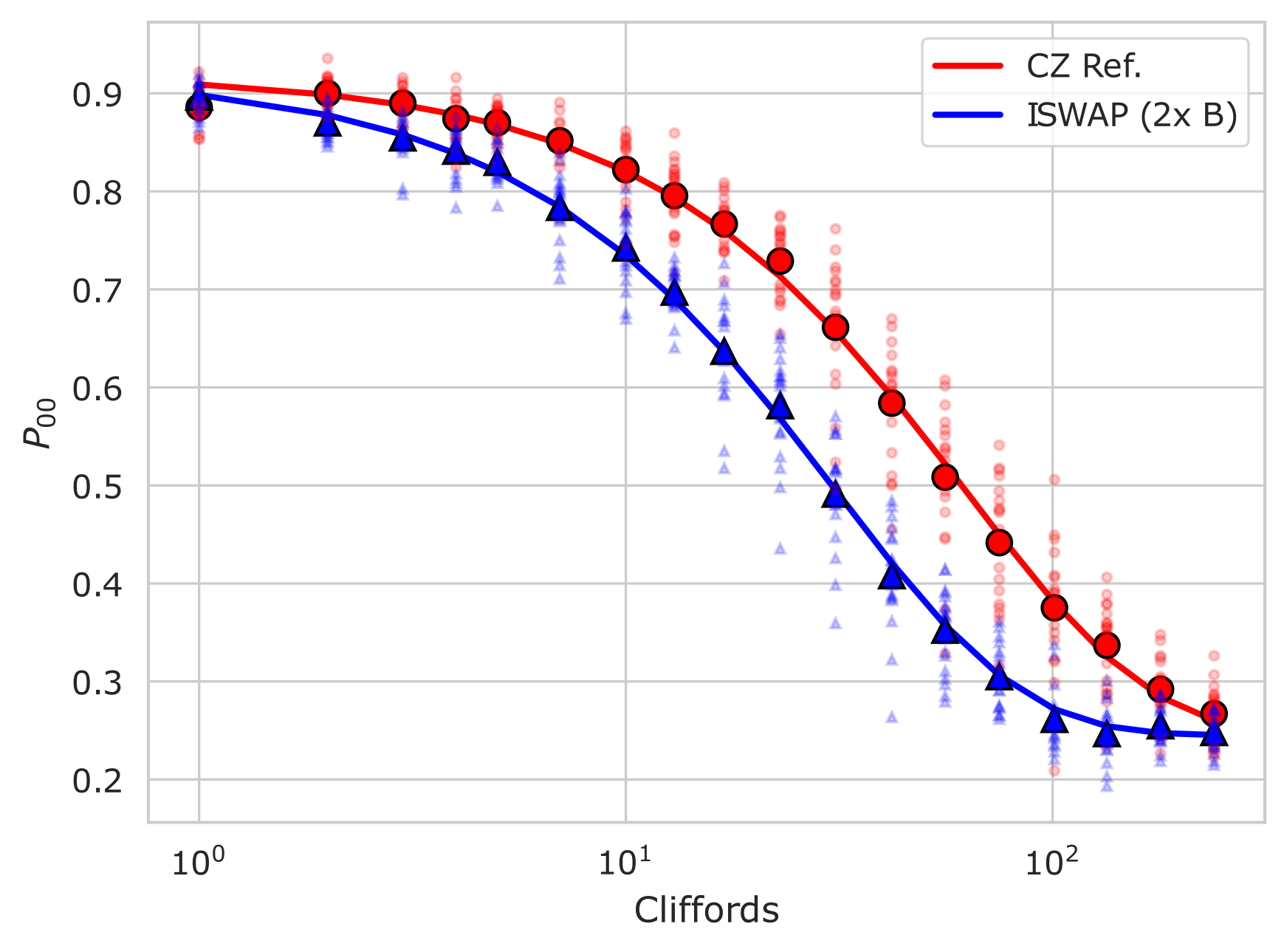}
\caption{Interleaved RB data for B gate. The interleaved element is an iSWAP gate composed of two B gates. The estimated EPG of the B gate is $0.00600(21)$, obtained by dividing the EPG of iSWAP by two.}
\label{bgate}
\end{figure}

B gate is the most efficient two-qubit gate in terms of compiling to SU(4) unitaries; any SU(4) can be compiled using single qubit gates and two B gates~\cite{Zhang2004}. The unitary of the B gate is given by $e^{i\pi (2XX+YY)/4}$. Due to the asymmetry between the two Ising interactions in the B gate unitary, it is not clear if there exits a native implementation of B gate. Here we realize B gate natively by combining $\text{iSWAP}^{3/4}$ and $\text{bSWAP}^{1/4}$ constructed using on-resonant drives described in section~\ref{sect:flicforq}. Notice that the bSWAP interaction can be generated by changing the single qubit rotations $Y_{\pm \pi/2}$ or the on-resonant drive from $+X$ to $-X$ in the iSWAP gate, however the frame tracking is reversed. The overall frame tracking for the B gate is given by the following
\begin{widetext}
\begin{align}
&\Qcircuit @C=0.5em @R=0.7em{
 &  \multigate{1}{\text{B}}  & \qw  & \raisebox{-2.2em}{$=$} & & \gate{Z(-\Delta t_1)} &  \gate{Y_{\pi/2}} & \gate{X_{3\Omega t_g/4}} & \gate{Y_{-\pi/2}} &    \gate{Z(\Delta t_2)}& \qw & \raisebox{-2.2em}{$+$} & & \gate{Z(\Delta t_2)} &  \gate{Y_{\pi/2}} & \gate{-X_{\Omega t_g/4}} & \gate{Y_{-\pi/2}} & \gate{Z(-\Delta t_3)} & \qw \\
 &  \ghost{\text{B}}   &\qw  & & & \qw & \qw & \qw & \qw   & \qw& \qw & & & \qw & \qw & \qw & \qw   & \qw & \qw} \nonumber \\ 
&\Qcircuit @C=0.5em @R=0.7em{
 & \raisebox{-2.2em}{$\Rightarrow$} & & \gate{Z(\phi_1)} & \gate{Y_{\pi/2}} & \gate{X_{3\Omega t_g/4}} & \gate{-X_{\Omega t_g/4}} & \gate{Y_{-\pi/2}} &    \gate{Z(\phi_2)} & \qw \\
 &  & & \gate{Z(\Delta t_\text{start}+ \phi_3)} & \qw & \qw & \qw & \qw   & \gate{Z(-\Delta t_\text{end} + \phi_4)} & \qw}
\end{align}
\end{widetext}
where we have used the properties $[XX+YY,ZI+IZ]=0 $ and $[XX-YY,ZI-IZ]=0$ to move the phases in between the $+X$ and $-X$ drives to the front and back of the gate. Interesting, frame tracking for the B gate is applied only on the idle qubit. Similar to $\sqrt{\text{iSWAP}}$ gate, we can go one step further and absorb the single qubit gates $Y_{\pm\pi/2}$ into single qubit unitaries in between B gates. Thus, the native implementation of B gate consists of just two on-resonant drives $+X$ and $-X$, with the $+X$ drive approximately three times as long as the $-X$ drive. There are four time-independent phases $\phi_{1,2,3,4}$ that need to be calibrated for the B gate, their calibration sequences are given in Appendix~\ref{sect:bcal}. We construct a iSWAP gate using two B gates and benchmark using IRB. The results are shown in Fig.~\ref{bgate}. The native B gate is 302 ns long, with the $+X$ and $-X$ drives being 210 ns and 92 ns respectively. The drives are flat-topped Gaussian pulses where rise and fall are $2\sigma$ long with $\sigma\approx 14.22$ ns. The measured EPG for B gate is comparable to iSWAP gate, this is expected since both gates are using the same mechanism and have comparable gate times. Similar to the $\sqrt{\text{iSWAP}}$ gate, whether the efficiency of B gate outweighs lack of virtual z-gates will be an important consideration for multi-qubit circuits. Finally, we remark that by combining on-resonant drives of opposite signs and CR, we can generate native two-qubit interactions of form $a XX + b YY + c ZZ$ with arbitrary coefficients. B gate is a special case with $c=0$, and $a=2b$, and SWAP gate is another with $a=b=c$. With the appropriate single qubit gates, our technique can generate arbitrary $SU(4)$ unitaries natively. We became aware of a related work recently~\cite{chen2023gate}.

\section{Application to circuits}

\begin{figure*}[tbh!]
\includegraphics[width=\textwidth]{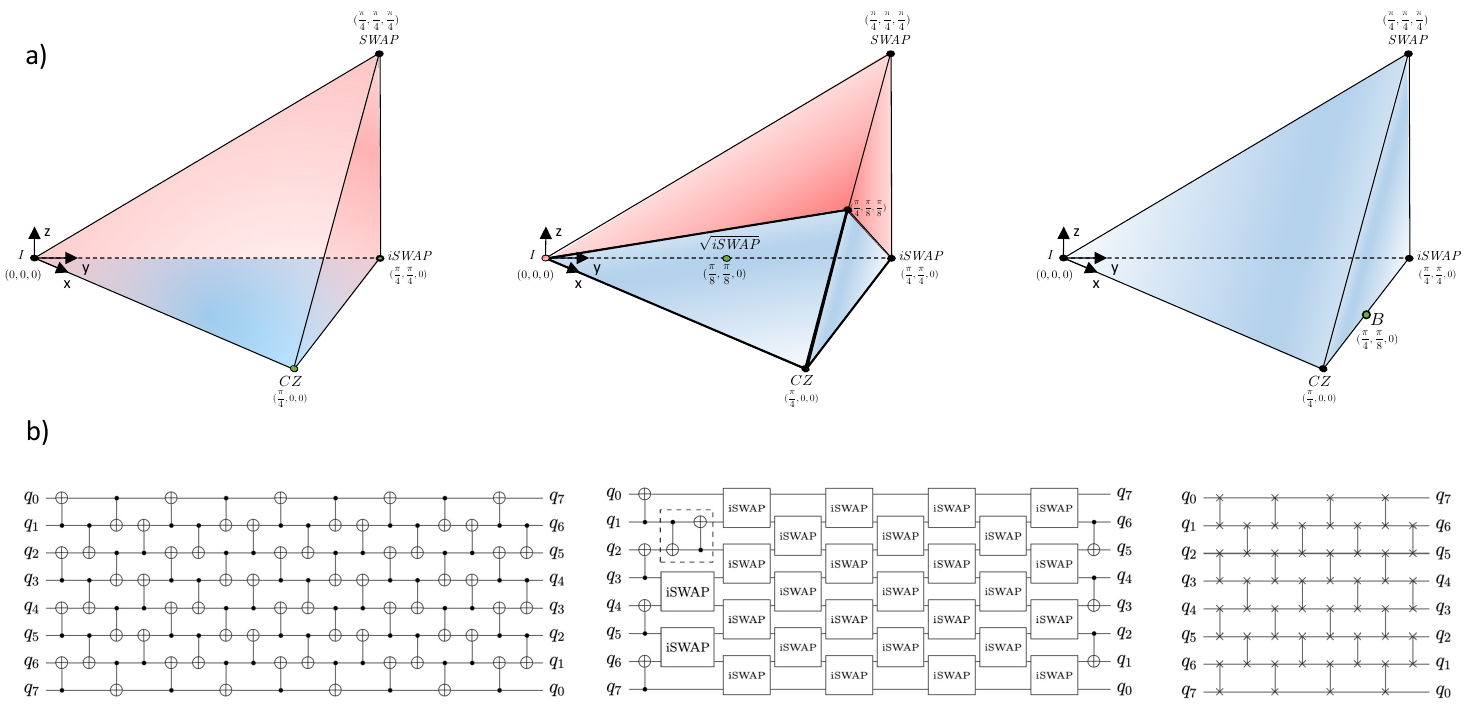}
\caption{Building circuits out of more expressive native gates. a) Regions of the Weyl chamber reachable by two (blue) or three (red) applications of the CZ gate or iSWAP gate (left),  the $\sqrt{\text{iSWAP}}$ gate (middle) and the B gate (right). b) Circuit compilation for the reversal of an $n$-qubit chain. The circuit requires depth $2n+2$ using only CZ (or CNOT) gates (left). Adding native iSWAP gates reduces the depth to $n+2$ (middle), and native SWAP gates require depth $n$ (right).}
\label{circuits}
\end{figure*}

The gates that we demonstrate in this work can create a richer and more expressive native instruction set on the hardware, which in turn can enable shorter and more efficient quantum circuits~\cite{peterson2022, huang2023}. In this section we discuss some potential applications of these gates in improving multi-qubit circuits.

As a first example, consider random two-qubit unitaries. These appear in quantum benchmarking protocols such as quantum volume, but also routinely during the compilation of circuits. In the latter setting, a compiler performs peephole optimization by collecting a sequence of single- and two-qubit gates, represents it as an instance of a two-qubit unitary, and attempts to resynthesize it with lower cost.  A common representation of the space of all two-qubit unitaries, $SU(4)$, is the Weyl chamber~\cite{zhang2003}. In this picture, any two-qubit unitary has a unique coordinate, up to local pre-/post-rotations. By repeated application of available native gates, one can traverse a path to reach a target unitary. The shorter this path, the higher the fidelity of the compiled circuit. Figure~\ref{circuits}(a) shows the region of unitaries reachable using various gates studies in this work. As we move from CZ to $\sqrt{\text{iSWAP}}$ to B gate, we gain a more expressive gateset that is able to reach more of the Weyl chamber with shorter paths.

We can consider other examples such as qubit permutations, an important task in limited-connectivity quantum processors. Figure~\ref{circuits}(b) shows possible circuits for the reversal of a chain of qubits on a linear-nearest-neighbor architecture, given different native gates. Using only CZ (or CNOT) gates, this circuit can be accomplished in depth $2n+2$~\cite{kutin2007computation}. However given access to iSWAP gates as well, the same circuit can be realized in depth $n+2$. Finally, native SWAPs can slightly improve the depth to $n$. While qubit reversals are important for example in performing quantum Fourier transforms or more general permutations~\cite{bapat2021}, our construction using native iSWAPs can extend to improve nearly all CNOT circuits on limited-connectivity hardware~\cite{de2021reducing}.

\section{Conclusion}

\begin{table}[b!]
\begin{ruledtabular}
	\begin{tabular}{c c c c}
	Gate &  EPG  &  Qubits used & Length (ns)\\
	\hline \vspace{0.0005\linewidth} \\
	iSWAP & $0.00648(21)$ &  [8, 9] & $398^*$\\
	SWAP & $0.00821(23)$ &   [8, 9] & $398^*$\\
	Stark iSWAP & $0.00409(27)$  &   [8, 9] & 192\\
	Stark bSWAP &   $0.00936(39)$ & [8, 12]  & 462 \\
	$\sqrt{\text{iSWAP}}$ & $0.00427(16)$ & [8, 9] & 149\\
	Stark $\sqrt{\text{iSWAP}}$ & $0.00294(15)$ & [8, 9] & 117\\
	B & $0.00600(21)$ & [8, 9] & 302 \\
	CZ & $0.00666(24)$ & [8, 9] & $512^*$
	\end{tabular}
\end{ruledtabular}
\caption{Summary of EPG for all the native two-qubit gates realized in this work. Gates with $*$ are benchmarked with single qubit rotations included.
}
\label{epg}
\end{table}

\begin{table*}[t!]
\begin{ruledtabular}
	\begin{tabular}{c c c c p{0.15\linewidth}}
	Gate &  Interaction  &  Unitary  & Resonance condition  & Frame tracking  \\
	\hline \vspace{0.001\linewidth} \\
	iSWAP (on-resonant) &  $XX+YY$  & $\left( \begin{array}{cccc} 1 & 0 & 0 & 0 \\ 0 & 0 & i & 0 \\ 0 & i & 0 & 0 \\ 0 & 0 & 0 & 1 \end{array}\right)$  & $\Omega=\Delta$  & $Z(\pm \Delta t_\text{end})$ behind driven/idle qubits 
	\vspace{0.1\linewidth} \\
	SWAP &   $XX+YY+ZZ$  & $\left( \begin{array}{cccc} 1 & 0 & 0 & 0 \\ 0 & 0 & 1 & 0 \\ 0 & 1 & 0 & 0 \\ 0 & 0 & 0 & 1 \end{array}\right)$ & $\Omega=\Delta+\omega_\text{s}$ & $Z(\pm \Delta t_\text{end})$ behind driven/idle qubits
	\vspace{0.1\linewidth} \\
	iSWAP (Stark) &   $XX+YY$  & $\left( \begin{array}{cccc} 1 & 0 & 0 & 0 \\ 0 & 0 & i & 0 \\ 0 & i & 0 & 0 \\ 0 & 0 & 0 & 1 \end{array}\right)$  & $\omega_\text{s}=\Delta$ & $Z(\pm \Delta t_\text{end})$ behind driven/idle qubits
	\vspace{0.1\linewidth} \\
	bSWAP (Stark)&   $XX-YY$  & $\left( \begin{array}{cccc} 0 & 0 & 0 & i \\ 0 & 1 & 0 & 0 \\ 0 & 0 & 1 & 0 \\ i & 0 & 0 & 0 \end{array}\right)$  & $\omega_1 + \omega_2 + \omega_\text{s}=2\omega_D$ & $Z(\omega_\text{s} t_\text{end})$ behind driven/idle qubits
	\vspace{0.1\linewidth} \\
	$\sqrt{\text{iSWAP}}$ (on-resonant) &   $XX+YY$ & $\left( \begin{array}{cccc} 1 & 0 & 0 & 0 \\ 0 & \frac{1}{\sqrt{2}} & \frac{i}{\sqrt{2}} & 0 \\ 0 & \frac{i}{\sqrt{2}} & \frac{1}{\sqrt{2}} & 0 \\ 0 & 0 & 0 & 1 \end{array}\right)$ & $\Omega=\Delta$ &  $Z(\pm \Delta t_\text{end})$ in front and back of driven qubit
	\vspace{0.1\linewidth} \\
	$\sqrt{\text{iSWAP}}$ (Stark) & $XX+YY$ & $\left( \begin{array}{cccc} 1 & 0 & 0 & 0 \\ 0 & \frac{1}{\sqrt{2}} & \frac{i}{\sqrt{2}} & 0 \\ 0 & \frac{i}{\sqrt{2}} & \frac{1}{\sqrt{2}} & 0 \\ 0 & 0 & 0 & 1 \end{array}\right)$ & $\omega_\text{s}=\Delta$ & $Z(\pm \Delta t_\text{end})$ in front and back of driven qubit 
	\vspace{0.1\linewidth} \\
	B Gate & $2XX+YY$ & $\left( \begin{array}{cccc} \cos\frac{\pi}{8} & 0 & 0 & i\sin\frac{\pi}{8} \\ 0 & \sin\frac{\pi}{8} & i\cos\frac{\pi}{8} & 0 \\ 0 & i\cos\frac{\pi}{8} & i\sin\frac{\pi}{8} & 0 \\ i\sin\frac{\pi}{8} & 0 & 0 & \cos\frac{\pi}{8} \end{array}\right)$ & $\Omega=\Delta$ & $Z(\mp \Delta t_\text{end})$ in front and back of idle qubit\\
	\end{tabular}
\end{ruledtabular}
\caption{Summary of native two-qubit gates realized in this work.}
\label{gate}
\end{table*}

In this manuscript we have implemented and benchmarked native iSWAP, SWAP, bSWAP, $\sqrt{\text{iSWAP}}$, and B gates in a fixed-frequency, fixed-coupling superconducting processor. A novel software time-dependent frame tracking technique is developed in Qiskit to correctly compile quantum circuits using these native two-qubit gates. We present in Table~\ref{gate} a summary of all the frame tracking used in each native two-qubit gate, and in Table~\ref{epg} the EPGs measured. We believe deploying these native gates in large scale superconducting processors could potentially benefit near term applications. The techniques developed for realizing these gates can be readily extended to qutrit systems~\cite{Goss2022}. Another important direction is to investigate the collision bounds for these gates~\cite{Hertzberg2021,zhang2022,morvan2022}; specifically, are they more flexible than CR gates in a fixed-coupling architecture? Preliminary work points to that these gates, like CR, might not work particularly well for very large detuning pairs, more studies are needed to understand why. Lastly, the frame tracking capabilities developed in Qiskit is general and can be readily applied to other architectures employing phase-sensitive two-qubit gates, such as the tunable architecture developed in~\cite{stehlik2021,zajac2021}. \\

\section{Acknowledgments}
The authors like to thank Oliver Dial, Karthik Siva, Matthias Steffen, Jiri Stehlik, and David Zajac for insightful discussions. This work was partially supported by the U.S. Department of Energy, Office of Science, National Quantum Information Science Research Centers, Co-design Center for Quantum Advantage (C2QA) under contract number DE-SC0012704.

\appendix

\begin{figure}[b!]
\includegraphics[width=0.45\textwidth]{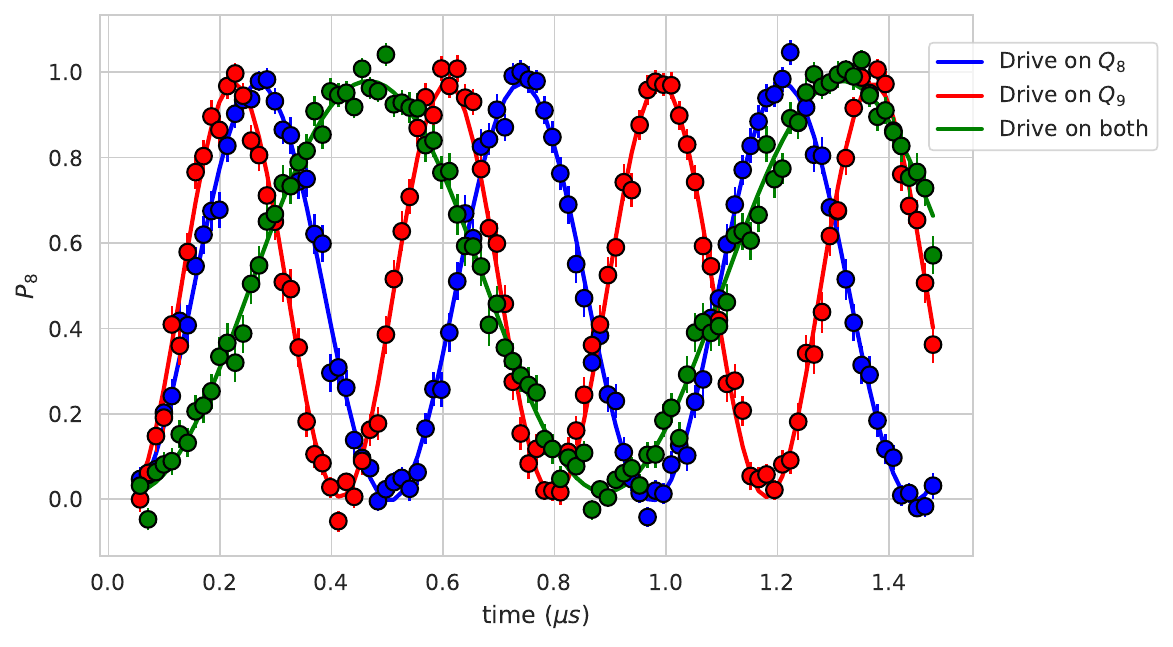}
\caption{$Q_8$ population as a function of time for the three driving scenarios shown in Fig.~\ref{flicforq}. The points with one standard deviation error bars are experimental data, and the curves are fits.}
\label{iswaposci}
\end{figure}
\section{iSWAP oscillations}
\label{sect:osci}
Here we show in Fig.~\ref{iswaposci} the oscillation dynamics for the on-resonant iSWAP drives. We fix the amplitude at the resonance point and sweep the length of the drive. We fit the oscillations to the equation $a \sin^2(2\pi f t + b) + c$, where $a$, $b$, $f$, and $c$ are fitting parameters with $f$ being the frequency. From the fits we find a rate of $1.062(2)$ MHz for driving on $Q_8$, $1.308(2)$ MHz for driving on $Q_9$, and $0.590(2)$ MHz for driving on both $Q_8$ and $Q_9$. Simultaneous driving achieves an iSWAP rate that is approximately half of that obtained from individual driving.

\section{CZ gate sequence} 
\label{sect:cz}
The pulse sequence for the CZ gate based on CR interaction is given by
\begin{align*}
&\Qcircuit @C=0.3em @R=0.7em{
 &  \multigate{1}{\text{CZ}}  & \qw  & \raisebox{-2.2em}{$=$} & & \qw &\gate{\Pi_{3\pi/4}}  & \multigate{1}{\text{CR}_{\pi/4}} &    \gate{X_{\pi}} & \multigate{1}{\text{CR}_{-\pi/4}} & \qw & \qw \\
 & \ghost{\text{CZ}}   &\qw  & & & \gate{Y_{\pi/2}}& \gate{X_{\pi/2}} & \ghost{\text{CR}_{\pi/4}}  & \qw & \ghost{\text{CR}_{-\pi/4}}& \gate{Y_{-\pi/2}} & \qw
}
\end{align*}
where $\Pi_{3\pi/4}=X_\pi \cos(3\pi/4) + Y_\pi \sin(3\pi/4)$ and $\text{CR}_{\pi/4}=ZX_{\pi/4}$ is the cross-resonance interaction.

\section{Phase Calibrations} 
Here we present the pulse sequences used to calibrate the time-independent phases required to realize each two-qubit gate. The first qubit is driven and second qubit is idle. For iSWAP, bSWAP, and SWAP gates there are two phases $\phi_{1,2}$, for $\sqrt{\text{iSWAP}}$ there are three phases $\phi_{1,2,3}$, and for B gate there are four phases $\phi_{1,2,3,4}$. All Z gates in the calibration sequences are implemented as virtual frame changes. All pulse sequences are Ramsey-type experiments which produces signal in the form $P\propto\frac{1+\cos(\phi-\phi_i)}{2}$ where $\phi$ is varied. The CNOT gates used in some calibration sequences are based on CR interaction. 

The calibration sequences for iSWAP, bSWAP, and SWAP are similar to single-qubit Ramsey experiments, except the second $\pi/2$ pulse is applied on the other qubit since the gate unitary switches the phases. The calibration sequences for $\sqrt{\text{iSWAP}}$ and B gates uses CNOT instead of $\pi/2$. It is a two-qubit version of the Ramsey experiment where we are trying to calibrate the relative phase in a Bell state such as $|00\rangle + e^{i\phi} |11\rangle$.

\subsection{iSWAP Phase Calibrations}
\label{sect:iswapcal}
\begin{align*}
\Qcircuit @C=0.5em @R=0.7em{
& \qw &  \multigate{1}{U_g} & \gate{Z(\Delta t_\text{end}+\phi_1)}  &    \gate{X_{-\pi/2}}  & \meter  \\
& \gate{Y_{\pi/2}} & \ghost{U_g}   & \qw &\qw  &\qw  
} 
\end{align*}
\begin{align*}
\Qcircuit @C=0.5em @R=0.7em{
& \gate{Y_{\pi/2}} &  \multigate{1}{U_g} &    \qw & \qw & \qw  \\
& \qw & \ghost{U_g} & \gate{Z(-\Delta t_\text{end}+\phi_2)}   & \gate{X_{-\pi/2}}  & \meter 
}
\end{align*} 

\subsection{SWAP Phase Calibrations } 
\label{sect:swapcal}
\begin{align*}
\Qcircuit @C=0.5em @R=0.7em{
& \qw &  \multigate{1}{U_g} & \gate{Z(\Delta t_\text{end}+\phi_1)} &    \gate{Y_{\pi/2}}  & \meter  \\
& \gate{Y_{\pi/2}} & \ghost{U_g}   & \qw &\qw  &\qw  
} 
\end{align*}
\begin{align*}
\Qcircuit @C=0.5em @R=0.7em{
& \gate{Y_{\pi/2}} &  \multigate{1}{U_g} &    \qw & \qw & \qw  \\
& \qw & \ghost{U_g}  & \gate{Z(-\Delta t_\text{end}+\phi_2)}  & \gate{Y_{\pi/2}}  & \meter 
}
\end{align*} 

\subsection{bSWAP Phase Calibrations } 
\label{sect:bswapcal}
\begin{align*}
\Qcircuit @C=0.5em @R=0.7em{
& \qw &  \multigate{1}{U_g} & \gate{Z(\omega_\text{s} t_\text{end}+\phi_1)}  &    \gate{X_{-\pi/2}}  & \meter  \\
& \gate{Y_{\pi/2}} & \ghost{U_g}   & \qw &\qw  &\qw  
} 
\end{align*}
\begin{align*}
\Qcircuit @C=0.5em @R=0.7em{
& \gate{Y_{\pi/2}} &  \multigate{1}{U_g} &    \qw & \qw & \qw  \\
& \qw & \ghost{U_g} & \gate{Z(\omega_\text{s} t_\text{end}+\phi_2)}   & \gate{X_{-\pi/2}}  & \meter 
}
\end{align*} 

\subsection{$\sqrt{\text{iSWAP}}$ Phase Calibrations } 
\label{sect:siswapcal}
\begin{align*}
\Qcircuit @C=0.5em @R=0.7em{
& \gate{Z(\Delta t_\text{start}-\phi_1)} &\targ &\gate{Z(-\Delta t_\text{start}+\phi_1)} &  \multigate{1}{U_g} &  \meter  \\
& \gate{X_{\pi/2}} &\ctrl{-1} &\qw & \ghost{U_g}   &\qw  
} 
\end{align*}
Let $\phi_{t,d}=\phi_2\pm\phi_3$, $\phi_{t,d}$ are calibrated according to the following sequences
\begin{align*}
\Qcircuit @C=0.5em @R=0.7em{
& \gate{X_{\pi}} &  \multigate{1}{U_g} &    \gate{Z(\Delta t_\text{end}+\phi_d)} & \targ & \qw &\qw  \\
& \qw & \ghost{U_g} & \qw   &\ctrl{-1} & \gate{X_{-\pi/2}}  & \meter 
}
\end{align*}
\begin{widetext}
\begin{align*}
\Qcircuit @C=0.5em @R=0.7em{
& \gate{Z(\Delta t_\text{start}-\phi_1)} &\targ &\gate{Z(-\Delta t_\text{start}+\phi_1)} &  \multigate{1}{U_g} &  \gate{Z(\Delta t_\text{end}+\phi_t)} & \targ & \qw &\qw  \\  
& \gate{X_{\pi/2}} &\ctrl{-1} &\qw & \ghost{U_g}   &\qw  &\ctrl{-1} & \gate{X_{\pi/2}}  & \meter 
} 
\end{align*}
\end{widetext} 

\subsection{B gate Phase Calibrations } 
\label{sect:bcal}
Let $\varphi_{t,d}=\phi_1\pm\phi_3$, $\varphi_{t,d}$ are calibrated according to the following sequences
\begin{align*}
\Qcircuit @C=0.5em @R=0.7em{
&\qw &\qw &\targ &\qw &  \multigate{1}{U_g} & \qw  \\
& \gate{Z(-\Delta t_\text{start}-\varphi_t)} &\gate{X_{-\pi/2}} &\ctrl{-1} &\gate{Z(\Delta t_\text{start}+\varphi_t)} & \ghost{U_g}   & \meter  
} 
\end{align*}
\begin{align*}
\Qcircuit @C=0.5em @R=0.7em{
&\qw &\gate{X_\pi} &\targ &\qw &  \multigate{1}{U_g} & \qw  \\
& \gate{Z(-\Delta t_\text{start}-\varphi_d)} &\gate{X_{-\pi/2}} &\ctrl{-1} &\gate{Z(\Delta t_\text{start}+\varphi_d)} & \ghost{U_g}   & \meter  
} 
\end{align*}
Let $\phi_{t,d}=\phi_2\pm\phi_4$, $\phi_{t,d}$ are calibrated according to the following sequences
\begin{align*}
\Qcircuit @C=0.5em @R=0.7em{
&  \multigate{1}{U_g} &    \qw & \targ & \qw &\qw  \\
& \ghost{U_g} & \gate{Z(-\Delta t_\text{end}+\phi_t)}   &\ctrl{-1} & \gate{X_{-\pi/2}}  & \meter 
}
\end{align*}
\begin{align*}
\Qcircuit @C=0.5em @R=0.7em{
& \qw &  \multigate{1}{U_g} &    \qw & \targ & \qw &\qw  \\
& \gate{X_{\pi}} & \ghost{U_g} & \gate{Z(-\Delta t_\text{end}+\phi_d)}   &\ctrl{-1} & \gate{X_{\pi/2}}  & \meter 
}
\end{align*} 

\bibliography{references}
\end{document}